\documentclass{aastex631} 

\usepackage{amsmath}

\shortauthors{Wang et al.}

\begin{document}

\title{Radiative Magnetohydrodynamic Simulation of the Confined Eruption of a Magnetic Flux Rope: Unveiling the Driving and Constraining Forces}

\correspondingauthor{Feng Chen}
\email{chenfeng@nju.edu.cn}

\author{Can Wang}
\affiliation{School of Astronomy and Space Science, Nanjing University, Nanjing 210023, China}
\affiliation{Key Laboratory for Modern Astronomy and Astrophysics (Nanjing University), Ministry of Education, Nanjing 210023, China}

\author[0000-0002-1963-5319]{Feng Chen}
\affiliation{School of Astronomy and Space Science, Nanjing University, Nanjing 210023, China}
\affiliation{Key Laboratory for Modern Astronomy and Astrophysics (Nanjing University), Ministry of Education, Nanjing 210023, China}

\author{Mingde Ding}
\affiliation{School of Astronomy and Space Science, Nanjing University, Nanjing 210023, China}
\affiliation{Key Laboratory for Modern Astronomy and Astrophysics (Nanjing University), Ministry of Education, Nanjing 210023, China}

\author{Zekun Lu}
\affiliation{School of Astronomy and Space Science, Nanjing University, Nanjing 210023, China}
\affiliation{Key Laboratory for Modern Astronomy and Astrophysics (Nanjing University), Ministry of Education, Nanjing 210023, China}

\begin{abstract}
We analyse the forces that control the dynamic evolution of a flux rope eruption in a three-dimensional (3D) radiative magnetohydrodynamic (RMHD) simulation. The confined eruption of the flux rope gives rise to a C8.5 flare. The flux rope rises slowly with an almost constant velocity of a few $\mathrm{km \ s^{-1}}$ in the early stage, when the gravity and Lorentz force are nearly counterbalanced. After the flux rope rises to the height at which the decay index of the external poloidal field satisfies the torus instability criterion, the significantly enhanced Lorentz force breaks the force balance and drives rapid acceleration of the flux rope. Fast magnetic reconnection is immediately induced within the current sheet under the erupting flux rope, which provides a strong positive feedback to the eruption. The eruption is eventually confined due to the tension force from the strong external toroidal field. Our results suggest that the gravity of plasma plays an important role in sustaining the quasi-static evolution of the pre-eruptive flux rope. The Lorentz force, which is contributed from both the ideal magnetohydrodynamic (MHD) instability and magnetic reconnection, dominates the dynamic evolution during the eruption process.
\end{abstract}

\section{Introduction} \label{sec:intro}

Flares and coronal mass ejections (CMEs) are usually regarded as the most energetic manifestations of solar eruptions: the former acts as the evidence of intense heating in the atmosphere \citep{Priest2002,Shibata2011,Janvier2015}, while the latter can carry massive plasma into the interplanetary space \citep{Webb2012,Green2018}. Although there was a debate on the causality between them in the early years, it has now become a general consensus that flares and CMEs are different representations of the same process, during which the magnetic energy is released violently and converted to the internal energy and kinetic energy of the plasma \citep{Zhang2001,Lin2004,Schmieder2015}.

Thanks to the launch of multiwavelength instruments and the development of high performance computers, the understanding on solar eruptions has been improved greatly in recent decades. Both observations and numerical simulations has confirmed that the magnetic flux rope, which is qualitatively identified as a group of magnetic field lines that twist around a common axis, acts as the key magnetic structure of solar eruptions \citep{Amari2000,Cheng2017,Liu2020}. The sigmoids in X-ray corona \citep{Rust1996,McKenzie2008} and the twsited hot channels in extreme ultraviolet (EUV) wavelength of high temperature \citep{Cheng2011,Zhang2012}, which are regarded as reliable evidence of flux ropes, have been captured right before or during the eruptions; the filaments/prominents, whose eruption is always related to flares and CMEs, may also indicate the existence flux ropes where the magnetic dips are able to hold cold and dense plasma in the corona \citep{Mackay2010,Parenti2014,Yan2015}. Meanwhile, data-constrained or data-driven simulations with the existence of flux ropes have restored many main features in observations, further substantiating the role of flux ropes in solar eruptions (see a comprehensive review by \citet{Jiang2022}).

Theories on eruption mechanisms have also advanced in recent years. Depending on whether magnetic dissipation is taken into account, the eruption models can be divided into two main categories. The first category is based on ideal magnetohydrodynamic (MHD) instabilities, for example, the helical kink instability \citep{Hood1981,Torok2004,Torok2005} which occurs when the twist number of the flux rope is high enough, and the torus instability \citep{Bateman1978,Kliem2006} which describes a catastrophic loss of balance between the downward strapping force from the external poloidal field and the upward hoop force as the a toroidal current ring rising. In models that resort to ideal MHD instabilities, there is usually a pre-eruption flux rope. In the second category, the eruption can be driven by magnetic reconnection \citep{Kusano2012} no matter whether there is a pre-existing flux rope. The reconnection can occur at a high-lying null point, which helps to remove the overlying field as indicated by the break-out model \citep{Antiochos1999}, or can occur in a low-lying site, such as in the tether-cutting model \citep{Moore2001} which can inject new flux and provide an upward Lorentz force to the erupting structure. 

However, it is still difficult to illustrate the dynamic evolution of the flux rope during its eruption in detail. Although present multiwavelength observations can capture the kinematic behaviors of the flux ropes, the in-situ detection of physical parameters in the corona, especially the magnetic field that dominates the dynamics of the flux rope in the low-$\beta$ circumstances, is still out of reach, and the identification of magnetic reconnection in a specific event is speculative. By comparison, sophisticated MHD simulations can achieve an accurate trace of the evolution of magnetic field and verify the role of ideal MHD instability and magnetic reconnection separately \citep{Aulanier2010,Jiang2021}; nevertheless, most of them are based on a simple preexisting magnetic structure (such as a bipole) and adopt simplified physical processes (in particular the radiative process), leading to a certain deviation between the simulated scenario and that on the real Sun. Although data-driven simulations use a series of observed photospheric magnetic field and velocity field as an input to the boundary, the results depend greatly on the governing equations, the option of how to input the observed parameters to the bottom boundary, as well as the treatment of background atmosphere \citep{toriumi2020}. Therefore, there is still a gap between the data-driven simulations and real observations.

On the other hand, eruptions of flux ropes can be either eruptive and confined. In the latter case, the eruption is accompanied by a flare but without a CME \citep{Ji2003,Liu2009,Joshi2014,Huang2020}. The causes that confine the eruption are also a debated topic. In the torus instability scenario, an eruption can be confined if the external field decays slower than the threshold of torus instability along the trajectory of the flux rope. For example, \citet{Guo2010} found a `saddle-like' profile of the decay index, which satisfies the condition of torus instability initially but then decreases below the threshold, this well explains why the eruption is initially triggered but finally confined. However, in some cases, the flux rope may fail to develop into a CME even if the decay index keeps large enough, which is termed as the failed-torus regime. In such cases, it was proposed that the external toroidal field may play a significant role by providing a substantial tension force \citep{Myers2015,Myers2017}, or that the non-axisymmetry of the flux rope can generate a downward Lorentz force and drag the flux rope down \citep{Zhong2021}. Additionally, the reconnection between the flux rope and the external field under the breakout-like configuration \citep{DeVore2008} or due to the rotation of the flux rope \citep{Zhou2019} can also provide possible explanations for confined eruptions.

In the past decades, the development of radiative magnetohydrodynamic (RMHD) simulations with sophisticated treatments on the active region evolution and energy transfer shed new lights on the understanding of the flux rope eruption. The magnetic field and plasma in such simulations evolve fully self-consistently with high fidelity, yielding synthetic images which are compatible with that of the real Sun. Compared with remote-sensing observations that only acquire line-of-sight integrated images, such realistic simulations can provide three-dimensional physical parameters within the key structures of the eruption, thus enabling us to explain in more detail the causes of the eruption and its possible confinement. In this work, we analyse the confined eruption of a flux rope in a comprehensive RMHD simulation, focusing on the driving and constraining forces before and during the eruption. 

The rest of the paper is organized as follows. The simulation setup and the selected data for analysis are given in Section \ref{sec:method}. In Section \ref{sec:result}, we analyse the relationship between the kinematic characteristics of the flux rope and the evolution of different forces exerting on the flux rope. Based on the results above, we discuss the contributions of torus instability and magnetic reconnection to the eruption in Section \ref{sec:discussion}, and the conclusions of this paper is given in Section \ref{sec:smr}.

\section{Method}\label{sec:method}
We analyze data from a previously done three-dimensional (3D) RMHD simulation \citep{Chen2022}. The simulation was conducted with the MPS/University of Chicago Radiative Magnetohydrodynamic (MURaM; \citep{Vogler2005,Rempel2017}) code. The MURaM code takes into account the radiative transfer in the optically thick regime, the optically thin radiative loss in the corona, the thermal conduction along magnetic field, and the ionization effects, which make the model a sufficient realism for direct comparisons with observations \citep{Cheung2019}. The energy loss in the chromosphere, transition region, and corona is balanced by heating provided by numerical resistive and viscous dissipation terms \citep{Rempel2017}.

The detailed setup of the specific simulation analyzed in this study has been presented in \citet{Chen2022}. The simulation contains a domain of $L_x \times L_y \times L_z = 196.608 \times 196.608 \times 122.88 \ \mathrm{Mm^3}$, spanning from the uppermost convection zone of about 9.6 Mm to over 100 Mm in the corona. The domain is resolved by 1024 $\times$ 1024 $\times$ 1920 grid points, yielding a spatial resolution of $\Delta x \times \Delta y \times \Delta z = 192 \times 192 \times 64 \ \mathrm{km^3}$. 

The initial condition of flux emergence stage is a snapshot of a quiet Sun simulation, where magneto-convection in the photosphere provides an upward energy flux that is dissipated in the upper atmosphere and maintains a million K hot corona. The quiet Sun simulation was evolved for a sufficiently long duration, such that the horizontally averaged temperature and density through the domain reach a dynamic equilibrium.

The formation and evolution of active regions is driven by the emergence of magnetic flux bundles generated in a global-scale solar convective dynamo \citep{Fan2014}. At the bottom boundary, a time series of the horizontal magnetic field and the full velocity vector extracted from the dynamo simulation is implemented as time-dependent boundary following the method described by \citet{Chen2017}. The entropy of upflows and the mean pressure are set to prescribed values. At the top boundary, the magnetic field in the ghost cells is a potential field calculated based on $B_z$ at the top of the computation domain; the vertical velocities in the first and second ghost cells are reduced to 50$\%$ and 25$\%$ of the values in the last two cells of the domain; the horizontal velocities and thermal variables are symmetric about the top boundary. The lateral boundaries of the domain are periodic for all variables.

Within 48 hours, the simulated active region gives rise to more than 50 flares of C class or above. We note again that the active region in the simulation originates from purely theoretically generated magnetic field. Therefore, we do not mean to model a particular active region or eruption event on the real Sun. Here, we only focus on a small cube (580 $\times$ 400 $\times$ 1620 grid points) containing the region of a C8.5 flare during a time period of approximate 900 s that starts from $\mathrm{21^h36^m55^s} \ (t_0)$, and all the moments mentioned in this paper is relative to $t_0$. The eruption starts at about 700 s, and the selected period nearly allows to cover the whole evolution of the pre-existing flux rope that gives rise to the flare.

\begin{figure*}
\centering
\includegraphics[width=\textwidth]{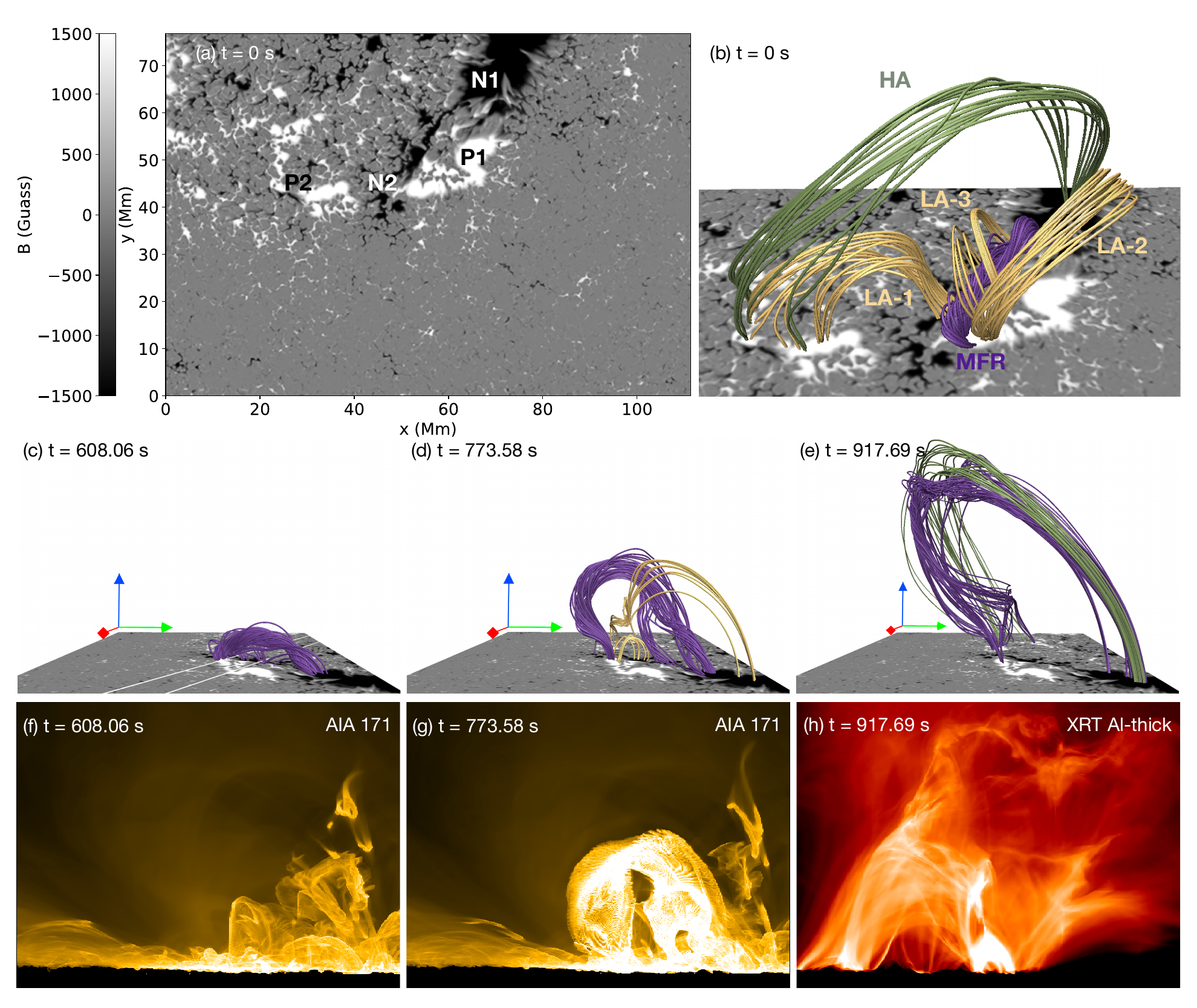} 
\caption{Overall evolution of the event. (a) Photospheric magnetogram of the selected cube at $t =$\ 0 s. P1 and N1 mark two strong and coherent sunspots, and P2 and N2 mark two weak and diffuse sunspots. (b) A zoomed view of main magnetic structures in the corona at $t =$\ 0 s. The twisted magnetic flux rope anchored in P1 and N1 is marked in purple. The golden lines outline three groups of magnetic arcades in the lower corona (LA-1, LA-2, and LA-3), while the higher arcades (HA) are marked in dark green. (c)-(e) Evolution of the flux rope before and during the confined eruption. The colors of magnetic field lines follow those in (b). The white lines in (c) mark the boundaries of the selected vertical planes within which we average the parameters to yield a 2D slice. The red, light green, and blue arrows indicate the $x$-, $y$-, and $z$-axis, respectively. The 3D visualization is produced by VAPOR \citep{Li2019}. (f)-(h) Synthetic AIA 171\ \AA\ and XRT Al-thick images from a side view during the same times as (c)-(e).}\label{fig:overall}
\end{figure*}

\section{Results} \label{sec:result}
\subsection{Kinematic Evolution of the Flux Rope} \label{subsec:kinematic}

An overview on the magnetic configuration and synthesized observables of the event is given in Figure \ref{fig:overall}. As shown in Figure \ref{fig:overall}(a), the photospheric magnetogram consists of two pairs of bipoles: the bipole P1-N1 appears to be strong and compact, while the other one (P2-N2) is relatively diffuse. Figure \ref{fig:overall}(b) provides a zoom-in view of the magnetic structures in the corona at $t = 0$ s. It is seen that a magnetic flux rope (MFR) has appeared before the flare onset, whose footpoints are anchored in the strong bipole P1-N1. In the lower corona, the magnetic arcades lie on the either side of the flux rope, denoted as LA-1 and LA-2, or straddle above it, denoted as LA-3; while in the higher corona, the arcades HA connect the main negative polarity N1 and the remote positive polarity P2. 

The temporal evolution of the flux rope and main observable signatures is shown in Figure \ref{fig:overall}(c)-(h). As the flux rope erupts, the lower arcades LA-1 and LA-2 reconnect within a current sheet under it, generating a group of cusp-like post-flare loops. The erupting flux rope appears as a bright arch in the synthetic Atmospheric Imaging Assembly (AIA; \citep{AIA}) 171\ \AA\ image, while the post-flare loops mainly show up in the synthetic X-Ray Telescope (XRT; \citep{XRT}) image. The detailed thermodynamic characteristics of the plasma associated with the flux rope eruption have been presented in \citet{Wang2022} (referred as Paper \uppercase\expandafter{\romannumeral1} hereafter). Note that while holding dense plasma, the flux rope is too cold before the eruption that there is no distinct observable counterpart of it in the synthetic EUV images during that time. Therefore, we follow the density structure, which can be identified as a blob significantly denser than its ambient, to describe the kinematic evolution of the flux rope. Moreover, we average the density over a series of vertical planes through the apex region of the flux rope, whose range is marked as the white lines on the photosphere in \ref{fig:overall}(c). This yields the density distribution on a two-dimensional (2D) vertical slice which is almost perpendicular to the axis of the flux rope. Most of the analysis mentioned below in this paper is based on the same slice.

\begin{figure*}
\centering
\includegraphics[width=\textwidth]{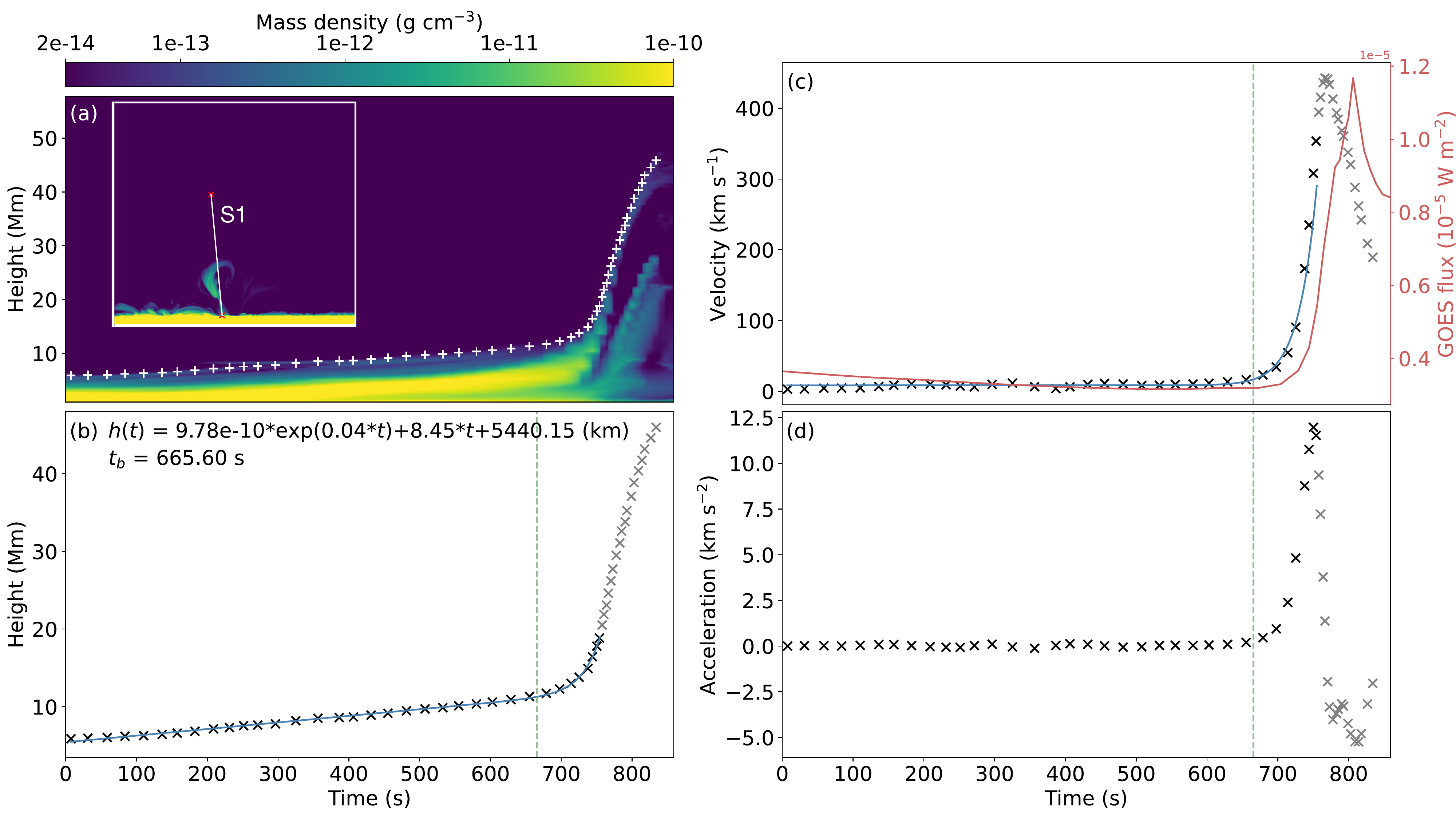} 
\caption{Kinematic evolution of the flux rope. (a) Height-time diagram of plasma density along slit S1, where the upper edge of the flux rope is marked by white crosses. The insert shows the position of S1 (white line) on the averaged 2D plasma density map. (b) The measured height of the flux rope fitted with a curve following the function marked at the top (b). (c) Velocity of the flux rope derived from the time derivative of the measured height. The red line shows the GOES 1–8\,\AA~flux. For comparison, the blue line shows the time derivative of the fitting function shown in (b). (d) Acceleration of the flux rope derived from the time derivative of the velocity. The black crosses in (b)-(d) indicate the data points used for curve fitting, while the excluded data are shown in gray. The green dashed lines in (b)-(d) indicate the observational onset time $t_b$.}\label{fig:kinematic}
\end{figure*}

Next, we choose a slit S1 following the trajectory of the flux rope eruption along which we can trace the main body of the flux rope on the 2D slice. Since the flux rope almost clings to the dense chromosphere at the beginning, and its shape changes greatly during the later stage of the eruption, it is difficult to identify its axis based on the density distribution. We finally adopt the upper edge of the flux rope along S1 as its location. Doing so may introduce small biases in the kinematic parameters but will not affect the main conclusions. The time-slit image of plasma density is presented in Figure \ref{fig:kinematic}(a), where the inserted panel shows the snapshot of the flux rope on the 2D slice, together with the slit S1 (the start and end points for height measurement are marked by red crosses). Since S1 is almost along the $z$-direction with a very small angle of about $5^{\circ}$ between them, the height variation of the flux rope can be approximately taken as its travel distance along the slit, as shown in Figure \ref{fig:kinematic}(a). Such an approximation provides a great convenience to measure the kinematic and dynamic parameters of the flux rope in the following analysis.

We measure the height of the flux rope at different time distance and plot the measured points in Figure \ref{fig:kinematic}(b). From the height-time measurement, we further derive the velocity and acceleration of the flux rope, which are the first and second order derivatives of the height with respect to time, as shown in Figure \ref{fig:kinematic}(c) and (d), respectively. The results show that, at the beginning, the flux rope rises slowly with a negligible acceleration for about 10 minutes. Then, it turns into the main-acceleration phase when both the velocity and acceleration increase rapidly. Afterwards, the acceleration tends to decrease and eventually becomes negative, implying a deceleration. Comparing the flux rope velocity with the synthetic Geostationary Operational Environmental Satellites (GOES) 1–8\,\AA~flux (Figure \ref{fig:kinematic}(b)), we find that they are basically synchronous during the eruption, except that the peak of the GOES flux is slightly delayed. In fact, a similar characteristic has also been revealed in observations for many confined eruptions \citep{Huang2020}.

According to \citet{Cheng2020}, the height-time variation of the flux rope during the slow-rise and the main-acceleration phases can be quantitatively described by the combination of a linear function and a exponential one:
\begin{equation}\label{eq1}
    h(t) = a{\rm e}^{bt}+ct+d.
\end{equation}
The time of turning point is defined as the moment when the nonlinear term begins to dominate the velocity:
\begin{equation}
    t_b = \frac{1}{b}{\rm ln}\frac{c}{ab}.
\end{equation}
The time $t_{b}$ can be regarded as the beginning of the impulsive acceleration, i.e., the observational onset time of flux rope eruption. In the height-time plot, we focus on the data points when the acceleration shows an increasing tendency, neglecting those when the acceleration is dropping down (gray crosses in Figure \ref{fig:kinematic}(b)-(d)). We then fit these data points using Equation (\ref{eq1}) and plot the result as the blue curve in Figure \ref{fig:kinematic}(b). The fitting yields a turning point at $t_b$ = 665.60 s, which is marked as vertical dashed lines in Figure \ref{fig:kinematic}(b)-(d). The derivative of the fitted function, i.e., the fitted velocity, matches the velocity derived from the measured height well, as shown in Figure \ref{fig:kinematic}(c).

\subsection{Force Balance in the Slow-rise Phase}\label{subsec:balance}
In the slow-rise phase, the acceleration of the flux rope is so small that it is roughly under a state of force balance. The forces in the simulation mainly consist of the following terms: the Lorentz force $\textbf{\textit{F}}_L = \textbf{\textit{j}} \times \textbf{\textit{B}} $, the gas pressure gradient $\textbf{\textit{F}}_p = - \nabla{P}$, and the gravity $\textbf{\textit{F}}_g = \rho \textbf{\textit{g}}$.

\begin{figure*}
\centering
\includegraphics[width=\textwidth]{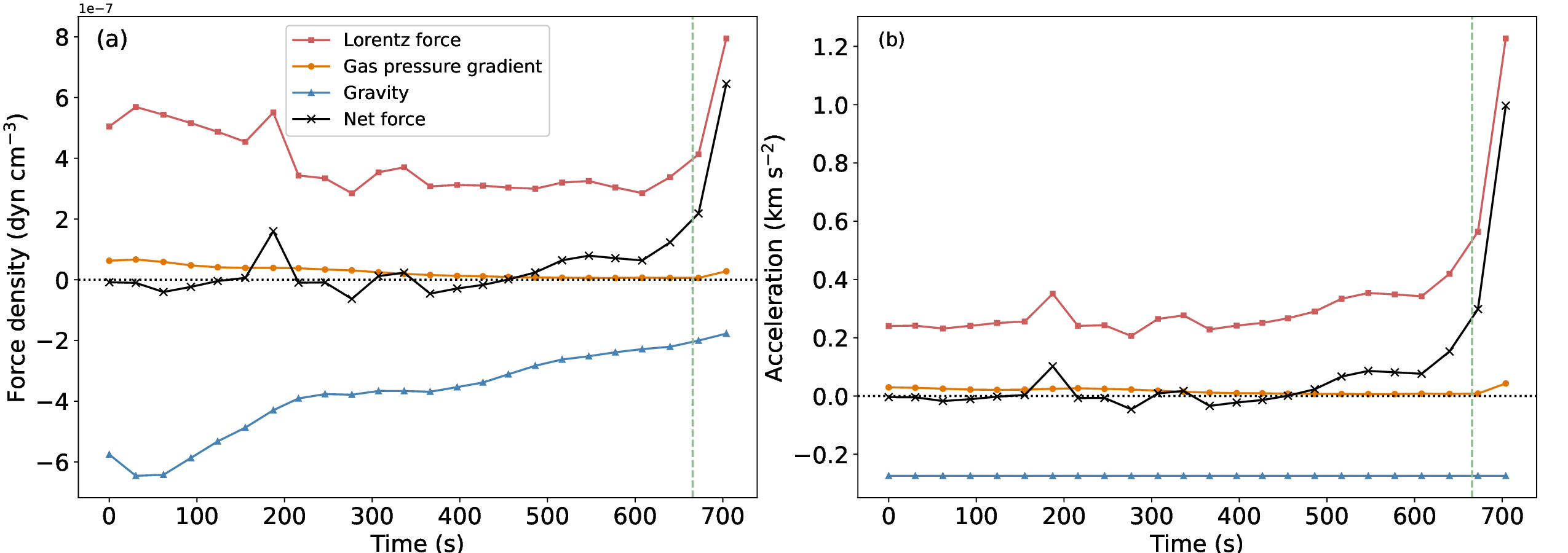}
\caption{Force and acceleration within the flux rope during the slow-rise phase. (a) Main force terms averaged over the flux rope region. The red, blue, and orange curves indicate the Lorentz force, gravity, and gas pressure gradient, respectively. The black curve marks the evolution of the net force. (b) The corresponding accelerations calculated by dividing each force by the mass of the flux rope, which are plotted in the same colors as that in (a). The green dashed lines indicate the observational onset time $t_b$.
}\label{fig:static}
\end{figure*}

During the slow-rise phase and the earlier period of acceleration phase, the flux rope keeps a regular morphology shaped with dense and cold plasma, and thus it is easy to determine the region it occupies based on the plasma density distribution. Figure \ref{fig:static}(a) plots the time variations of the Lorentz force, gravity, pressure gradient, and the net force averaged over the flux rope region on the 2D slice from 0 s to about 700 s. Figure \ref{fig:static}(b) gives the corresponding accelerations, if we divide each force by the mass the flux rope carries over the same region. 

We notice that the gravity plays an important role in balancing the upward Lorentz force, resulting in a rough force balance condition since the pressure gradient is one order of magnitude smaller. Note that the evolution of acceleration calculated directly from the force is consistent with that derived from the kinematic measurement: in the first 500 s or so, the net acceleration is trivial and only slightly fluctuates around zero; it then rises above zero and shows an increasing tendency for a short time period before the onset time $t_b$, indicating that the flux rope may have erupted for some time before the observationally determined onset time. After the onset time, both the net force and the corresponding acceleration grow violently, which will be discussed in the following section.

\subsection{Dynamic Evolution during the Eruption}
\subsubsection{Onset of the Torus Instability}\label{subsec:torus}

Substantial studies have indicated that torus instability can effectively initiate the eruption of flux ropes \citep{Aulanier2010,Inoue2018,Kliem2021}, which occurs when the decay index
\begin{equation}
    n(z) = -\frac{d\,({\rm ln}B_{\rm p,ex})}{d\,({\rm ln}z)}
\end{equation}
is larger than a critical value. Here, $B_{\rm p,ex}$ is defined as the poloidal component of the external magnetic field. For events with a simple bipole-like magnetic configuration, $B_{\rm p,ex}$ is usually approximated by the transverse component of the potential field, which can be extrapolated from the pre-eruption photospheric magnetogram. Here, we select a horizontal layer where the average optical depth is unity as the photosphere, and apply the $B_z$ component on this layer at a snapshot just before $t_b$ as the bottom boundary condition for extrapolation. The potential field in the corona is then extrapolated by the condition of vanishing $B_z$ at infinity and periodicity in the horizontal direction.

Since the magnetic field distribution in our simulation is far more complex than a bipole configuration, in order to more reasonably calculate the decay index, we try to decompose the horizontal potential field into the poloidal field and the toroidal field. Considering that the current averaged over the flux rope region on the 2D slice is dominated by its toroidal component, we roughly regard the direction of the averaged current as the alignment of local flux rope axis. The external poloidal field is then defined as

\begin{equation}
 \begin{aligned}
    B_{\rm p,ex} &= \textbf{\textit{B}}_{\rm h,po} \cdot (\bold{i}_j \times \bold{i}_z) \\
    &= \frac{B_{x,{\rm po}}j_y-B_{y,{\rm po}}j_x}{\sqrt{j_x^2+j_y^2}},\label{bs}
 \end{aligned}
\end{equation}
while the toroidal one is calculated as:
\begin{equation}
 \begin{aligned}
    B_{\rm t,ex} = - \bold{i}_j \cdot \textbf{\textit{B}}_{\rm h,po}, \label{bg}
 \end{aligned}
\end{equation}
where the negative sign is added considering that the toroidal current of the flux rope is nearly antiparallel with the toroidal magnetic field of it. In Equations (4) and (5), $\textbf{\textit{B}}_{\rm h,po}$ is the horizontal component of the potential field, and $\bold{i}_j$ and $\bold{i}_z$ are unit vectors along the toroidal current and the $z$-direction, respectively.

\begin{figure*}
\centering
\includegraphics[width=\textwidth]{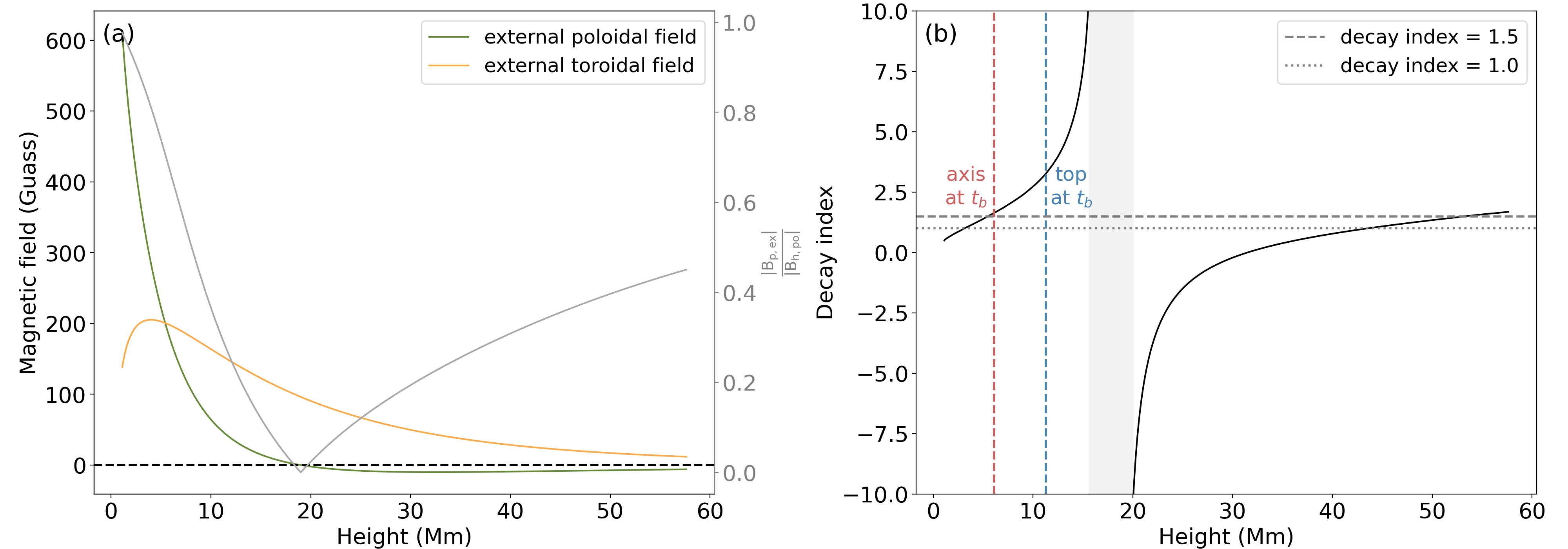} 
\caption{The magnetic field and decay index along the trajectory of the flux rope. (a) Distribution of the external poloidal field (green curve) and the external toroidal field (yellow curve) versus height along the eruption trajectory S1 on a sampled slice. The ratio of the external poloidal field to the horizontal component of the potential field is shown as the gray curve. (b) Distribution of the mean decay index versus height, which is averaged over the sampled positions of the flux rope. The gray dotted and dashed lines indicate the value of decay index at 1.0 and 1.5, respectively. The positions of the upper edge and the axis of the flux rope at $t_b$ are marked with blue and red vertical dashed lines, respectively.
}\label{fig:TI}
\end{figure*}

We calculate the quantities $B_{\rm p,ex}$, $B_{\rm t,ex}$, and the decay index along the eruption trajectory S1 for a series of sampled positions along the axis of the flux rope. Figure \ref{fig:TI}(a) shows the distribution of magnetic field versus height measured for a specific sampled position. We find that within a height range of about 20 Mm from the photosphere, the poloidal field $B_{\rm p,ex}$ induces a downward Lorentz force. However, it decreases rapidly and even changes sign above 20 Mm. The toroidal field $B_{\rm t,ex}$ first increases and then decreases with height, keeping parallel to the toroidal field of the flux rope.
The mean decay index profile, which is averaged over the sampled positions of the flux rope, is shown in Figure \ref{fig:TI}(b). Note that owing to the reversal of sign of $B_{\rm p,ex}$, the averaged decay index jumps from more than 10 to less than -10 within a small height range (covered by the gray shadow in the figure). We also plot in the figure two critical values of decay index, 1.0 and 1.5, which are usually regarded as thresholds of torus instability for direct and semicircular current tubes, respectively.

To judge whether torus instability contributes to the eruption, we focus on the decay index at the position of flux rope axis at $t_b$. The height of the flux rope axis is obtained by subtracting the radius of the flux rope from the height of the upper edge. Figure \ref{fig:TI}(b) indicates that at the beginning of the impulsive acceleration, the decay index at the flux rope axis reaches 1.5, suggesting that the torus instability may contribute to the eruption.

\subsubsection{Magnetic Reconnection under the Flux Rope}\label{subsec:reconnection}

The appearance of the cusp-like post-flare loops, as well as the current sheet under the flux rope shown in Paper \uppercase\expandafter{\romannumeral1}, strongly hint the occurrence of magnetic reconnection here, which may also break the force balance and play a role in driving the eruption. We further focus on the evolution of different force terms during the eruption and try to find out the possible contribution from magnetic reconnection. Different from the behavior in slow-rise phase, the flux rope undergoes an obvious deformation after eruption. On the one hand, the flux rope shows an anisotropic expansion; on the other hand, it reconnects with the ambient field. Meanwhile, the thermodynamic properties of plasma hosted in the flux rope change greatly. The rapid changes of magnetic field and plasma make it difficult to determine the boundary of the flux rope and calculate the average forces within it during the eruption. Therefore, we analyze the distribution of forces along S1 instead, which can still capture the main dynamic features of the flux rope.

\begin{figure*}
\centering
\includegraphics[width=\textwidth]{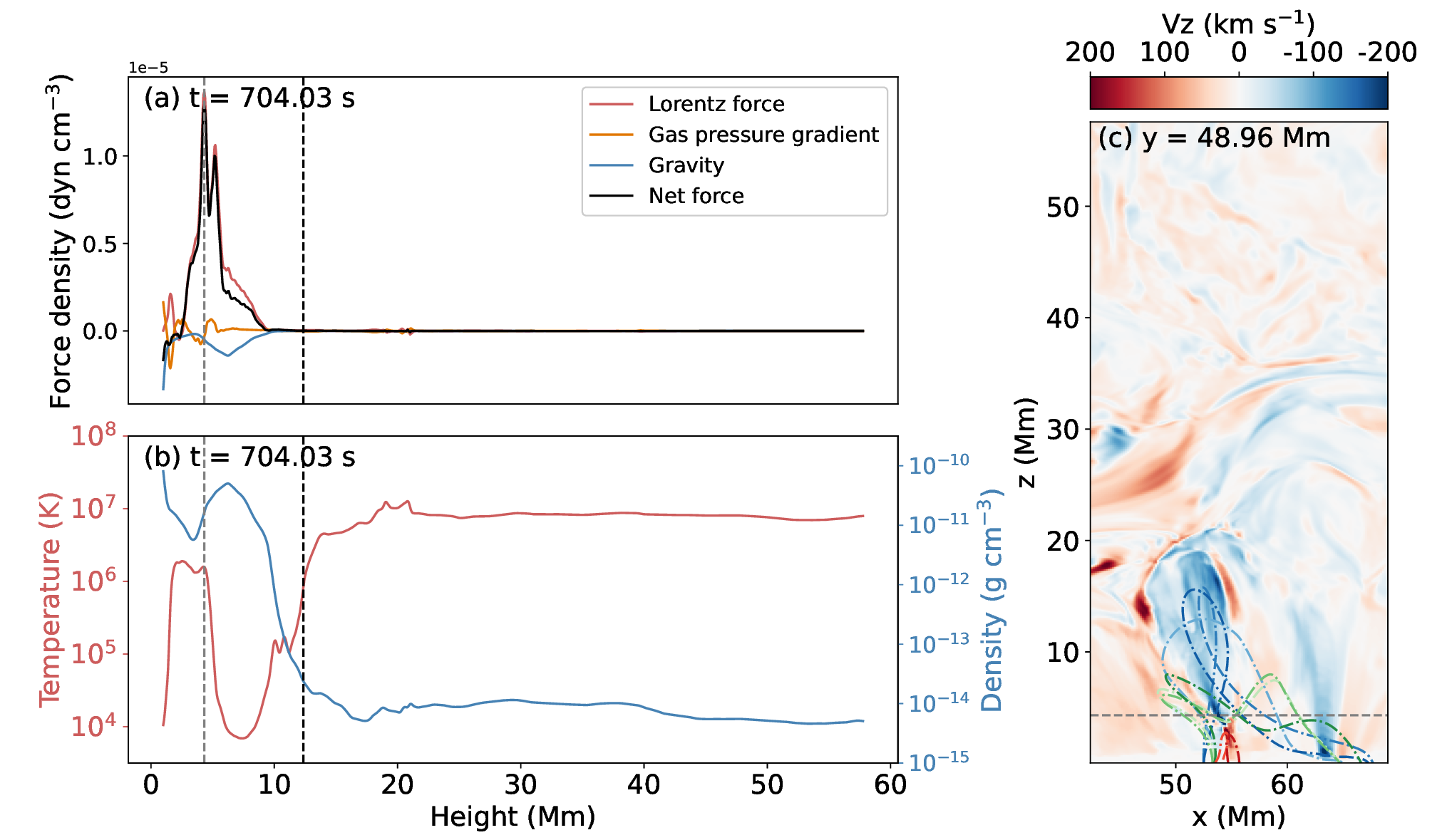}
\caption{Dynamic behaviours and reconnection characteristics during the eruption. (a) Distribution of the Lorentz force (red), the gas pressure gradient (orange), the gravity (blue), and the net force (black) along S1 at $t$ = 704.03 s. (b) Distribution of temperature (red) and plasma density (blue) along S1 at $t$ = 704.03 s. (c) Map of $v_z$ on a $x$-$z$ plane located at $y = $\ 48.96 Mm at $t$ = 704.03 s. Magnetic field lines corresponding to the flux rope (blue), the newly reconnected arcades (green), and the post-flare loops (red) are projected in the plane. The black dashed lines in (a) and (b) mark the upper boundary of the flux rope calculated by Equations (1), and the gray dashed lines in (a)-(c) mark the position of the maximum Lorentz force.}\label{fig:reconnection}
\end{figure*}

Figure \ref{fig:reconnection}(a) shows the distribution of the Lorentz force, the gravity, the gas pressure gradient, and the net force along S1 at $t$ = 704.03 s, which is approximately 40 s after $t_b$. The plasma properties along S1 at the same time are shown in Figure \ref{fig:reconnection}(b), where we can approximately distinguish the location of the flux rope, which is relatively denser and cooler than the higher corona. Comparing Figure \ref{fig:reconnection}(a) with \ref{fig:reconnection}(b), we find that within the flux rope, the curve of the net force almost overlaps with that of the Lorentz force, indicating the dynamic evolution of the flux rope is dominated by the Lorentz force during this period. Moreover, while keeping a significantly high density, the plasma at the left wing of the force peak is two orders of magnitude hotter than its surroundings, which hints an efficient heating consistent with the scenario of magnetic reconnection. We then plot the $v_z$ map on an $x$-$z$ plane (located at $y = $\ 48.96 Mm) in Figure \ref{fig:reconnection}(c), where evident bidirectional outflows with velocities of more than 200 $\rm km\  s^{-1}$ can be detected near the height of the force peak. For a more intuitive view, we project the magnetic filed lines on the $x$-$z$ plane and overplot some selected lines on both sides of the outflows. The blue dot-dashed lines are field lines within the flux rope. It is seen that the upper parts of these lines possess a closed tear-drop shape, which in fact represents a twisted morphology in three dimensions. Below this, there appear arc-shaped lines (in red) that outline the post-flare loops. The field lines (in green) sandwiched between the flux rope and the flare loops correspond to the newly reconnected field lines from LA-1 and LA-2 mentioned in Figure \ref{fig:overall}(b), which show a shape concave upward with the dip standing above the post-flare loops. All of the evidence above suggests that there is magnetic reconnection occurring during the eruption, which helps to drive the eruption by providing an upward Lorentz force.

\subsection{Confinement of the Eruption}\label{subsec:failed}

\begin{figure*}
\centering
\includegraphics[width=0.7\textwidth]{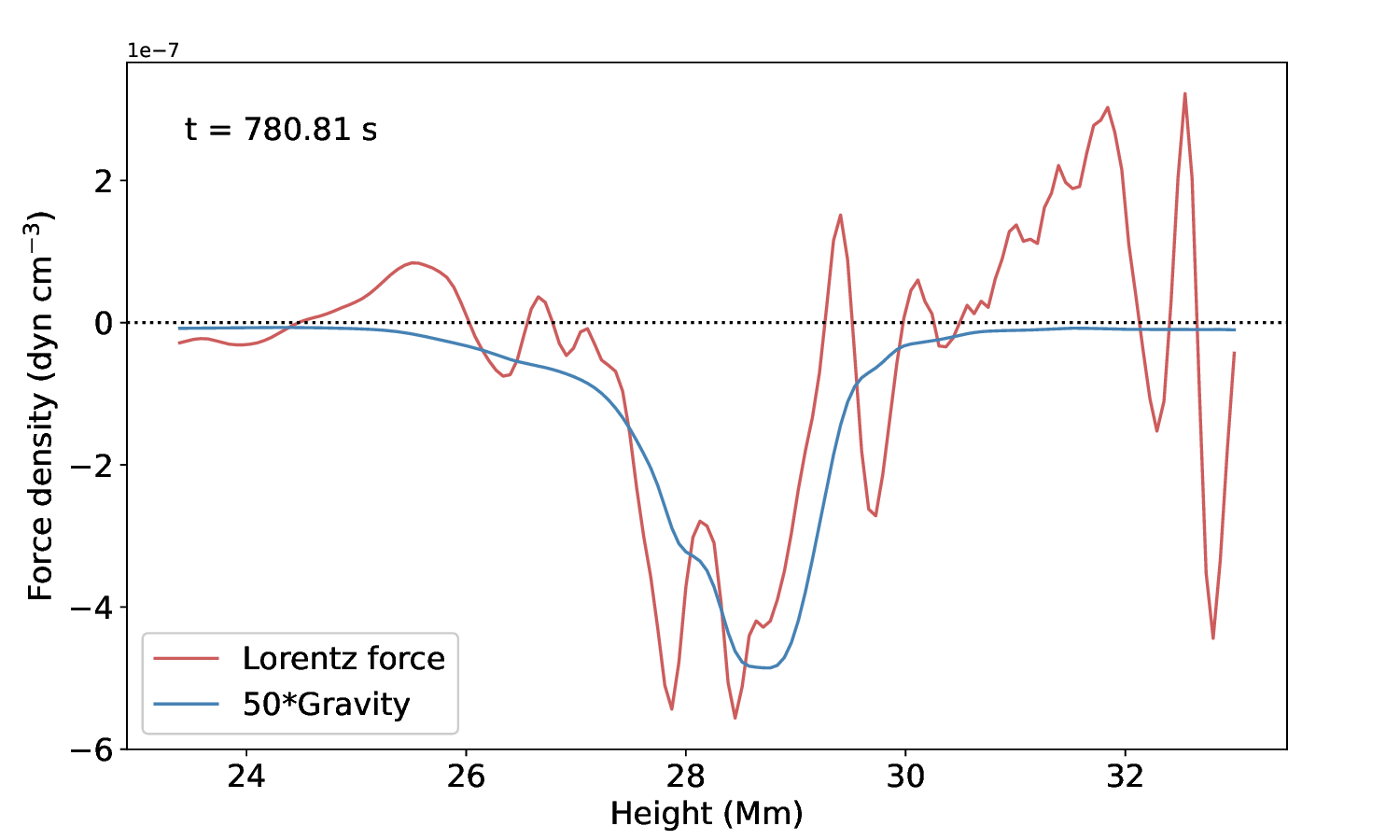}
\caption{Distribution of forces across the flux rope along S1 at $t$ = 780.81 s when the flux rope is decelerating. The red line presents the Lorentz force, and the gravity multiplied by 50 is shown by the blue line.}\label{fig:failed}
\end{figure*}

The measured kinematic characteristic shown in Figure \ref{fig:kinematic}(d) reveals that during the deceleration phase, the magnitude of the deceleration of the flux rope is much greater than the gravitational acceleration on the solar surface, suggesting that there are other factors contributing to the confinement of the eruption other than the gravity. We further concentrate on a small height range of the flux rope along S1 to check the contribution of the Lorentz force, which plays the most important role in driving the eruption during the main-acceleration phase. The distribution of the Lorentz force at $t$ = 780.81 s is shown as the red curve in Figure \ref{fig:failed}, together with that of the gravity, from which we can locate the position of the flux rope that hosts dense plasma. We find that both the gravity and the Lorentz force are directed downward within most part of the flux rope, but the Lorentz force obviously dominates the dynamic behavior of the flux rope. As a result, it is clear that the confinement of the eruption occurs when the Lorentz force becomes negative.

Generally speaking, if the external poloidal field decays too slowly to trigger the torus instability, the downward strapping force will finally defeat the upward hoop force and stabilize the flux rope. However, the decay index in our case exceeds the theoretical threshold for the onset of torus instability, making this event in the failed-torus regime. Note that although the decay index becomes negative above about 20 Mm, it has no effect on the confinement of the eruption, which will be discussed in Section \ref{subsec:profile}.

Focusing on Figure \ref{fig:overall}(b), we notice that as the flux rope rises, the angle between the axis of the flux rope and the external magnetic field (higher arcades HA) seems to be much smaller than that in the lower corona, indicating that the external magnetic field becomes more toroidal in that height range. Both observations \citep{Chen2021} and experiments \citep{Myers2015,Myers2017} have suggested that the toroidal field that provides a downward tension force when interacting with the poloidal current of the flux rope can effectively contribute to the confinement of the eruption; in simulation conducted by \citet{Kliem2014}, the toroidal field also plays a key role to sustain a stable configuration when it is strong enough. According to Figure \ref{fig:failed}, the flux rope reaches a height of approximate 25-30 Mm when it begins to decelerate. During such a height range, the ratio of the external poloidal field $|B_{\rm p,ex}|$ to the total horizontal potential field $|B_{\rm h,po}|$ is less than 0.2 (see the gray line in Figure \ref{fig:TI}(a)), strongly suggesting the dynamic evolution of the flux rope depends mainly on the external toroidal field instead of the poloidal one. Meanwhile, the upward Lorentz force resulting from the magnetic reconnection below the flux rope gradually fades out and can no longer power the eruption. 
. Therefore, the external toroidal field, which acts as a confining cage as shown in Figure \ref{fig:overall}(e), can effectively prevent the eruption of the flux rope.

There are other factors that may contribute to the confinement of the eruption. For example, as indicated by \citet{Zhong2021}, the non-axisymmetry of the flux rope can induce a downward Lorentz force and hence confine the eruption. Moreover, the reconnection between the flux rope and the external field can erode the flux rope and weaken the toroidal current within it, making the hoop force of the flux rope fall below the strapping force from the external field \citep{DeVore2008}. In our simulation, the flux rope expands anisotropically during the eruption; meanwhile, the reconnection between the flux rope and ambient field also occurs, as indicated by the corrugated shape of magnetic field lines near the current shell, which wrap around the flux rope with plasma heated to more that 10 MK (see Figure 4 in Paper \uppercase\expandafter{\romannumeral1}). As a result, we do not exclude their roles in confining the eruption of the flux rope.

In a word, besides the hoop force and strapping force that exert on the flux rope as in the ideal torus instability model, the tension force results from the toroidal field can take great effect in the flux rope dynamics. Other component of the Lorentz force that results from the nonaxisymmetry of the flux rope, as well as reconfiguration of the flux rope and the ambient field due to magnetic reconnection, may also be valued. As a result, the flux rope is finally constrained although the profile of the decay index still implies a successful eruption.

\section{Discussion} \label{sec:discussion}
\begin{figure*}
\centering
\includegraphics[width=0.8\textwidth]{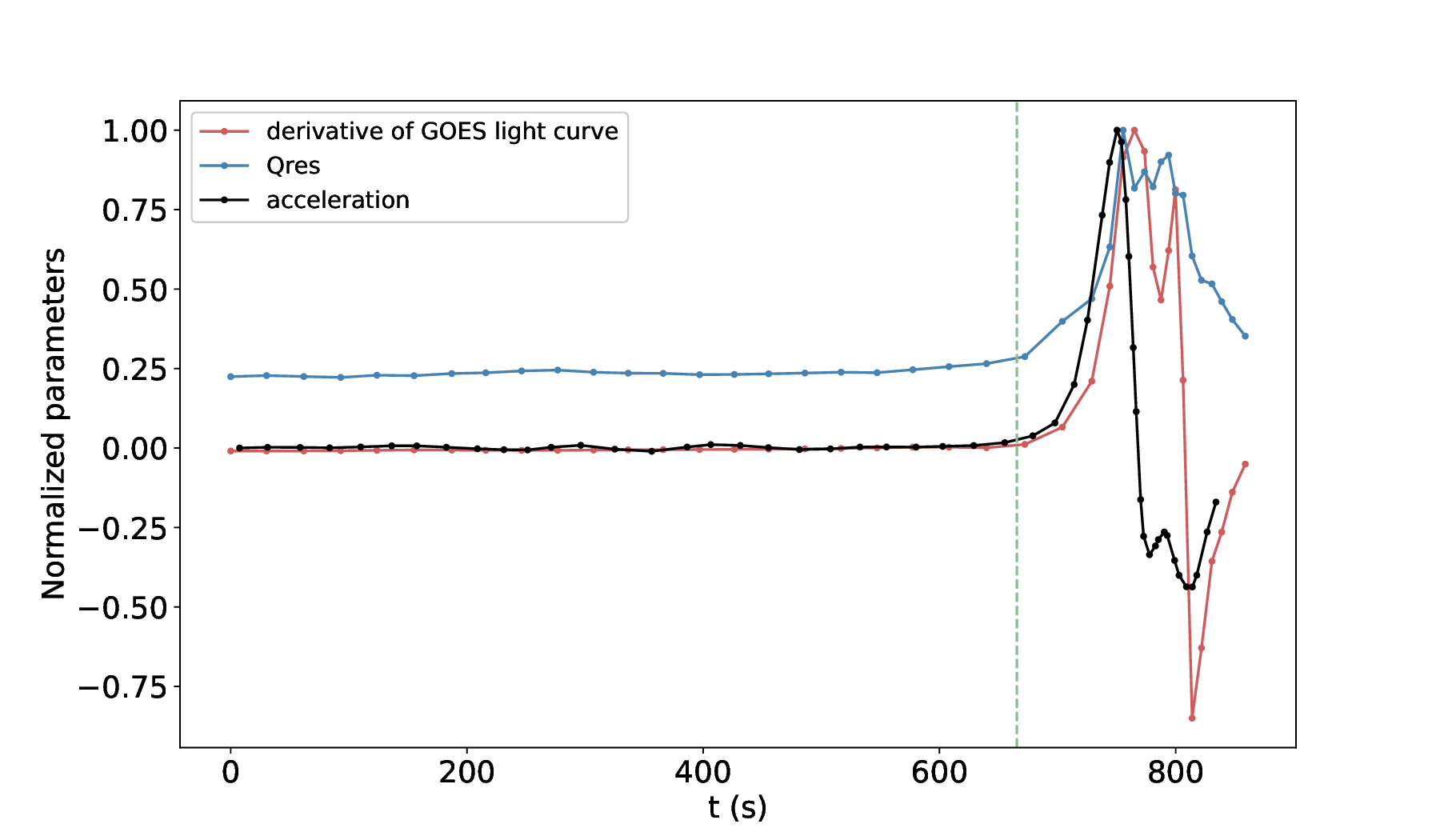}
\caption{Acceleration of the flux rope compared with the evolution of reconnection efficiency. The numerical resistive dissipation term $Q_{\rm res}$ integrated within a region containing the current sheet under the flux rope (blue) and the derivative of synthetic GOES flux (red) are used as proxies of the reconnection efficiency. The acceleration is the same as that in Figure \ref{fig:kinematic}(d).}\label{fig:q}
\end{figure*}

\begin{figure}
\centering
\includegraphics[width=0.5\textwidth]{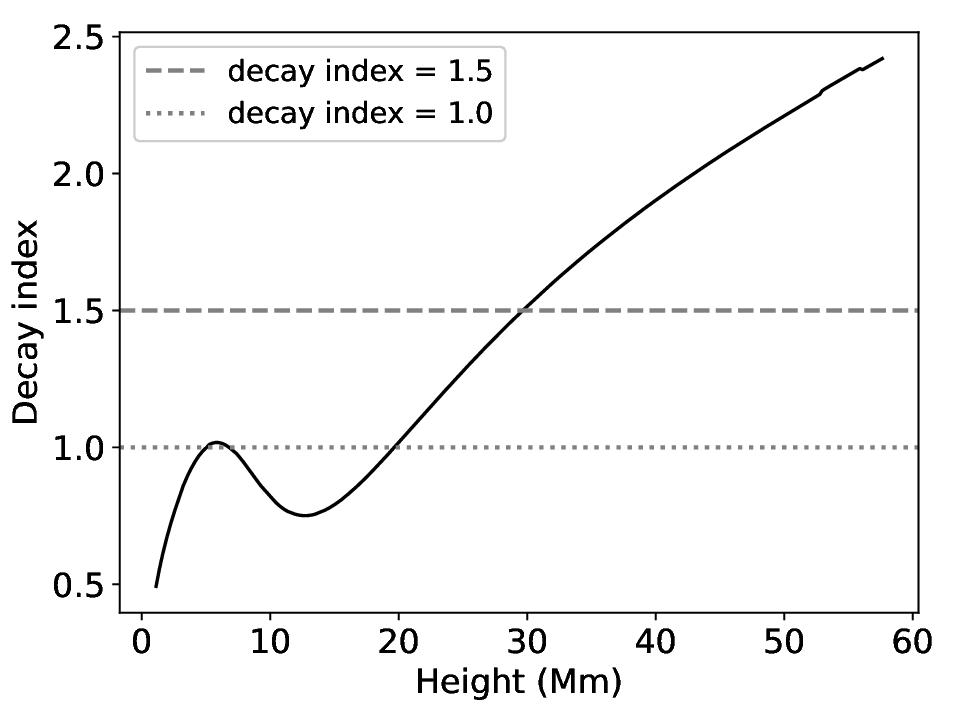}
\caption{Decay index calculated with the full horizontal component of the potential field, which includes both the poloidal component and the toroidal component. The gray dotted and dashed lines indicate the value of decay index at 1 and 1.5, respectively.}\label{fig:btotal}
\end{figure}

\subsection{Details on the Evolution during the Slow-rise Phase}\label{subsec:discussion_forcebalance}
Since the plasma-$\beta$ in the corona is expected to be very small, the evolution of coronal plasma is thought to be dominated by the magnetic field. Zero-$\beta$ assumption is widely adopted in theoretical models and numerical simulations, where the gravity is always balanced with the gas pressure gradient and the evolution of the flux rope is determined solely by the Lorentz force. However, the force balance of plasma-hosting flux ropes needs to comprehensively account for the contribution by plasma forces, in order to more accurately describe the dynamics of flux ropes. 

The gravity acts as the key factor to maintain the quasi-static evolution of the flux rope during its slow-rise phase in this event. Such a scenario is similar to that in \citet{Fan2017,Fan2018} and \citet{Jenkins2019}, where the mass-loading or mass-draining can actually influence the onset of eruption. Our result suggests that the consideration of gravity may help to obtain a more realistic evolution of the magnetic field before and in the early stage of the eruption, especially in events with a filament/prominence containing plasma that is much denser than surrounding corona; moreover, the sophisticated energy balance considered in simulations, which is important for more accurate modeling of the plasma properties, allows for more realistic evolution on both dynamics and thermodynamics and provides synthetic observables that is of great consistence with the observations, as also shown by \citet{Fan2022}.

\begin{figure*}
\centering
\includegraphics[width=\textwidth]{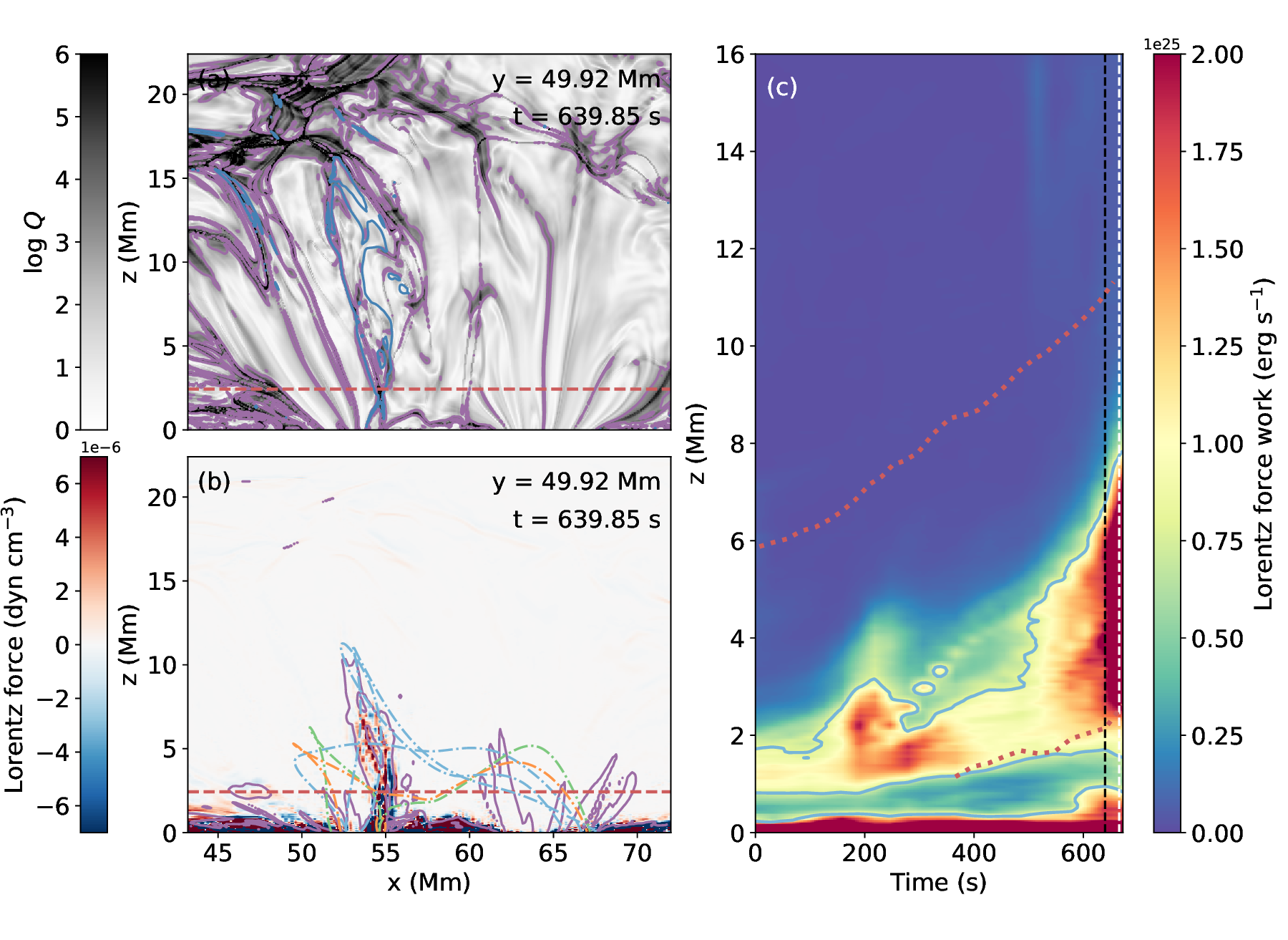} 
\caption{Distribution of Lorentz force and Lorentz force work in the slow-rise phase. (a) Distribution of quashing factor $Q$ on an $x$-$z$ plane located at $y$ = 49.92 Mm at $t$ = 639.85 s. The purple and blue contours show the quashing factor with a level of log$\ Q$ = 3 and the twist number with a level of $T_w$ = -1.5, respectively. (b) Distribution of the $z$-component of Lorentz force on the same plane as that in (a) at $t$ = 639.85 s. The purple contour shows the current density with a level of $j = \rm{2 \times 10^{-7}\ Guass\  cm^{-1}}$. The magnetic field lines before and after magnetic reconnection as well as that selected within the flux rope are projected in the plane as green, orange, and blue dot-dashed lines, respectively. The horizontal red dashed lines in both (a) and (b) mark the height of the lower boundary of the flux rope on this plane. (c) Height-time diagram of the Lorentz force work (color bar) integrated within each horizontal layer with the height of one pixel. The blue contour marks the value of $P$ = $\rm 8 \times 10^{24}\ erg\ s^{-1}$. The higher and lower red dotted lines mark the averaged lower and upper boundaries of the flux rope, respectively. The time of the snapshot shown in (a) and (b) is indicated by the black vertical dashed line, while the onset time $t_b$ is indicated by the white vertical dashed line.
}\label{fig:power}
\end{figure*}

The period before $t_b$ that appears to be a uniform stage in the kinematic evolution can be, however, roughly divided into two sub-stages when focusing on the evolution of forces, as shown in Figure \ref{fig:static}. At the beginning, the gravity and Lorentz force averaged within the flux rope show similar decreasing trends, suggesting that the expansion of the flux rope may, at least partly, contribute to the slow rise of its upper edge detected in Section \ref{subsec:kinematic} during this period. Meanwhile, the state of force balance is well maintained, as indicated by the net force and acceleration that fluctuate slightly around zero. After $t \approx$ 500 s or so, the averaged gravity continues to decrease due to the on-going expansion of the flux rope; however, the Lorentz force gradually prevails over the gravity and drives the flux rope to deviate from the force balance regime. 

To understand whether the enhancement of Lorentz force is a large-scale variation within the flux rope as described by torus instability, or comes from local effect such as magnetic reconnection under the flux rope, we investigate the evolution of its spatial distribution during the slow-rise phase. We locate the lower boundary of the flux rope with quashing factor $Q$ \citep{Priest1995,Titov2002} and twist number $T_w$ \citep{Berger2006} with the method provided by \citet{Zhang2022}. Figure \ref{fig:power}(a) shows the distribution of $Q$ on an $x$-$z$ plane located at $y$ = 49.92 Mm at $t$ = 639.85 s, where the purple contour and blue contour mark regions with relatively higher $Q$ and $T_w$, respectively. The X-shaped feature in both $Q$ and $T_w$ contours located at $x \approx$ 55 Mm and $z \approx$ 2.5 Mm is regarded as the lower boundary of the flux rope here, whose height is marked by the horizontal red dashed line. Figure \ref{fig:power}(b) shows the distribution of the $z$-component of Lorentz force on the same plane as that in (a), where the purple contour marks the region with relatively strong currents.

The height-time diagram shown in Figure \ref{fig:power}(c) illustrates the evolution of the Lorentz force work $P = \textbf{\textit{v}}_z \cdot \textbf{\textit{F}}_{Lz}$ integrated within each horizontal layer with the height of one pixel, where we only take regions with both upward force and upward velocity into account \footnote{Taking both the upward and downward part into account will not put significant effect on the overall conclusion; however, since the reconnection site is located at lower atmosphere where the distribution of Lorentz force is chaotic, including everything will make the Lorentz force contributed by magnetic reconnection partially canceled out after integration.}. The instance of Figure \ref{fig:power}(a)-(b) and the observational eruption onset time $t_b$ are marked by black and white vertical dashed lines, respectively. At each snapshot, we calculate the $Q$ and $T_w$ maps on a series of planes at different locations to locate the averaged lower boundary of the flux rope, and by repeating this process at different snapshots we finally obtain the evolution of the averaged lower boundary, which is shown as the lower red dotted line in Figure \ref{fig:power}(c), while the higher red dotted line approximately marks the upper boundary of the flux rope as we has measured in Section \ref{subsec:kinematic}.

We find that at $t \approx$ 500 s, the filling fraction of large Lorentz force work within the flux rope increases from approximately 17$\%$ to 37$\%$, which hints the development of large-scale upward Lorentz force within the flux rope that drives its acceleration. The evolution during this period is consistent with the scenario of the torus instability, in which the flux rope is driven by hoop force applied on its main body. Meanwhile, we project magnetic filed lines on the $x$-$z$ plane and overplot some selected lines corresponding to different magnetic structure in Figure \ref{fig:power}(b), where the green and orange dot-dashed lines are related to magnetic field lines before and after magnetic reconnection that occurs near the bottom of the flux rope. However, there is no evident local enhancement of Lorentz force work at the bottom of the flux rope, which indicates that although the magnetic reconnection can work to change the magnetic configuration and provide flux to the flux rope, it may not provide efficient upward tension force to accelerate the flux rope directly. Our result suggests that during the later stage of the slow-rise phase, the main driving force is the large-scale Lorentz force that appears as torus instability begins to develop.

\subsection{Identifying the Possible Trigger of the Eruption}\label{subsec:onset}
Previous works have suggested that either torus instability or fast reconnection can efficiently drive the eruption. The initiation and development of ideal MHD instability and magnetic reconnection are very likely to couple tightly with each other \citep{Liu2021}. Therefore, the key question of which factor works as the initial trigger remains a puzzle. 

A quantitative estimate on the criteria of ideal MHD instability under more realistic circumstances is non-trivial. For example, for a large aspect-ratio current tube undergoing a self-similar expansion, torus instability is usually thought to be satisfied theoretically when the decay index at its axis reaches 1 (for straight current tubes) or 1.5 (for semicircular ones). However, taking the factors such as the morphology of the flux rope \citep{demoulin2010}, the line-tied effect \citep{isenberg2007,filippov2021}, the photospheric evolution \citep{zuccarello2015}, and the role of gravity \citep{Jenkins2019} into consideration, there may be a critical range rather than a universal critical value for the onset of torus instability. Meanwhile, due to the lack of in-situ observations, the local decay index at the position of the flux rope relies greatly on the way to choose its observational indicators \citep{zuccarello2016,sarkar2019,rees2020}. In addition, the decay index at the eruption height can also be affected by the definition of the onset time. As a result, it may be not convincing enough to identify the trigger mechanism only based on a single value of decay index. By carefully performing a group of relaxation calculations, \citet{Aulanier2010} excluded the role of bald-patch reconnection, tether-cutting reconnection, and flux disappearance in initiating the eruption and finally credited the onset of eruption in their simulation to torus instability.

The analysis in Section \ref{subsec:balance} suggests that the flux rope has begun to accelerate slowly but steadily at $t \approx $ 500 s, which may manifest as the earliest signature of the eruption \citep{Cheng2020}. As shown in Figure \ref{fig:power}(c), the large-scale Lorentz force has begun to develop before $t_b$, indicating that although the decay index has not reached the theoretical threshold of 1.5, the torus instability is potential to make sense for triggering the eruption at that time. The magnetic morphology also implies that the actual threshold of torus instability may be less than 1.5 in this event: the axis of the flux rope is almost flat, which corresponds to a lower-lying current cube with a smaller fractional number of the partial torus. Under such a circumstance, the flux rope is more unstable and owes to a smaller threshold of torus instability even though taking the gravity into consideration \citep{Cargill1994,Olmedo2010}.

As shown by \citet{Zhao2017} and \citet{Jiang2021}, in events that are difficult to meet the requirement of torus instability, the eruption is only triggered when fast reconnection occurs. The reconnection efficiency is usually described by the reconnection rate within the current sheet, which can be calculated with the reconnection electric field \citep{Yokoyama2001} or external Alfv$\rm{\acute{e}}$n Mach number \citep{Zhao2017}. In observations, the motion of flare ribbons \citep{Qiu2002} is often used to estimate the reconnection electric field; the hard X-ray flux as well as the derivative of soft X-ray emission \citep{Neupert1968} are also related to the real-time reconnection efficiency. Here, we use the numerical resistive dissipation term $Q_{\rm res}$, which is calculated based on monotonicity change of the magnetic field and measures the dissipation of magnetic energy \citep{Rempel2017}, as the indicator of magnetic reconnection. Figure \ref{fig:q} shows the evolution of $Q_{\rm res}$ integrated within a box that contains the current sheet under the flux rope. It indicates that, in this event, the fast reconnection occurs later than the onset of the eruption, as demonstrated by the small delay of the impulsive increases of both the $Q_{\rm res}$ and the derivative of GOES soft X-ray curve relative to the measured acceleration. We also inspect the temperature and velocity field under the flux rope and find no signature of impulsive heating or acceleration of local plasma in this region before $t \approx $ 670 s.

To summarize, our results suggest that the torus instability acts as the initial trigger of the eruption. In the early stage of the event, the flux rope rises slowly in the lower atmosphere. Slow magnetic reconnection may occur under the flux rope, which contributes to its slow rise but does not trigger an eruption \citep{Aulanier2010}. Once the torus instability develops sufficiently, the flux rope turns into the main-accelerate phase. The local magnetic pressure decreases abruptly after the flux rope moves above, which facilitates the formation of a thin current sheet where fast reconnection occurs and establishes a positive feedback to the eruption of the flux rope.

\begin{figure*}
\centering
\includegraphics[width = \textwidth]{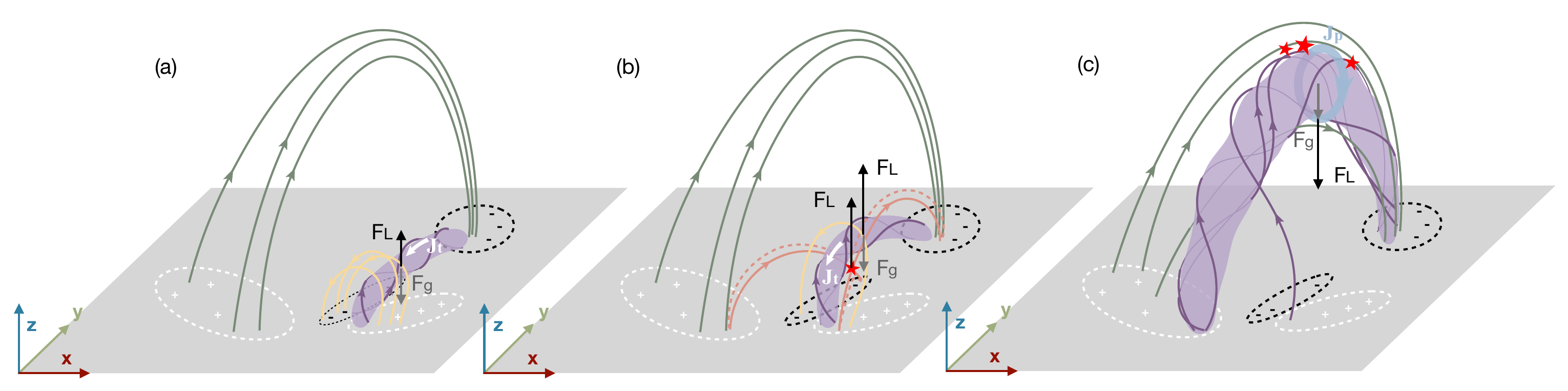}
\caption{
Cartoon of the confined eruption describing the evolution of the flux rope before and during the eruption. (a) The slow-rise phase of the flux rope. (b) Eruption of the flux rope. (c) Confinement of the eruption. The gray plane marks the photosphere, where the white and black dotted circles correspond to the positive and negative magnetic poles, respectively. The purple tube surrounded by a group of purple lines marks the main body of the magnetic flux rope. The yellow and orange lines mark the magnetic arcades in the lower corona, while the higher arcades are drawn as dark green lines. The white arrow marks the toroidal current $J_{\rm t}$ within the flux rope, while the light blue one surrounding the flux rope in (c) indicates the poloidal current $J_{\rm p}$ of it. The red pentagrams indicate the reconnection site. The dark red, light green, and dark blue arrows at the bottom left corner refer to the $x$-, $y$-, and $z$-axes, respectively. See Section \ref{sec:smr} for more details. }\label{fig:cartoon}
\end{figure*}

\subsection{Profile of Decay Index in the Confined Eruption}\label{subsec:profile}

Over the past decades, observations have revealed that instead of increasing monotonically versus height, the decay index sometimes shows a saddle-like distribution with a smaller value at the saddle dip \citep{Guo2010}.  Moreover, in some events there is a negative saddle where the decay index is even smaller than zero \citep{Filippov2020}. Such a saddle-like profile is considered to be a reasonable explanation for some confined eruptions that meet the requirement of torus instability in a lower height, where the eruption is initially triggered but finally gets constrained in the saddle region if the flux rope has not accumulated a large enough momentum. \citep{liu2018}.

To explore the origin of the the saddle-like profiles, \citet{Luo2022} employed dipoles of different strengths or positions and found that the saddle-like profiles may be associated with specific magnetic skeletons, such as hyperbolic flux tube (HFT) or null point. The decay index in our work shows a profile with a negative saddle, which is, however, given rise by the sign reversal of the external poloidal field $B_{\rm p,ex}$, instead of a null point. Note that the magnetic field above the reverse point, with the external poloidal field in the opposite direction to that below it, can even contribute to the eruption of the flux rope than hold it back.

Like in most events, the relative orientation of the external magnetic field to the flux rope cannot be precisely determined due to the complexity of the magnetic field and dynamic evolution of the flux rope during its eruption. Thus, we also calculate the decay index using the full transverse field for the same region and at the same time as that in Section \ref{subsec:torus}. The result is shown in Figure \ref{fig:btotal}. Interestingly, this yields a more typical saddle-like profile, which can explain the confined eruption but seems to fail to meet the threshold for torus instability at a lower height. Such a difference between the two calculations comes from the complex configuration of the magnetic field, which contains a considerable toroidal component. Note that although the toroidal field can provide a tension force to constrain the eruption, it had better be excluded from the analysis of torus instability which is related with the strapping force from the poloidal field.

\section{Conclusion} \label{sec:smr}
We analyze the confined eruption of a magnetic flux rope in a 3D RMHD simulation, focusing on its dynamic evolution before and during the eruption, especially the driving and constraining forces. We summarize the main conclusions as cartoons in Figure \ref{fig:cartoon}, which can be described as follows:

\begin{enumerate}

\item The kinematic evolution of the flux rope can be roughly divided into a slow-rise phase, a main-acceleration phase, and a confined phase. The kinematic features during the slow-rise and the main-acceleration phases can be described by a fitting function consisting of a linear term and an exponential term, from which we can derive the observational onset time.

\item The flux rope undergoes an approximate force-balance condition during its slow-rise phase, when the gravity plays an important role in balancing the upward Lorentz force and maintaining the quasi-static evolution of the flux rope (Figure \ref{fig:cartoon}(a)). 

\item The eruption begins earlier than the observational onset time and is indicated by the steady accumulation of Lorentz force before $t_b$. Torus instability provides large-scale upward Lorentz force to drive the acceleration of flux rope during the latter stage of the slow-rise phase and works as the most likely trigger of the eruption.

\item Fast magnetic reconnection occurs within the current sheet under the flux rope once the torus instability develops, leading to an abrupt increase of the Lorentz force and establishing a positive feedback to the eruption. The dynamics of the flux rope is then driven by the Lorentz force from both torus instability and magnetic reconnection (Figure \ref{fig:cartoon}(b)).

\item The tension force from the strong external toroidal field is most likely the factor to confine the eruption. The effects of the non-axisymmetry of the flux rope, as well as the magnetic reconnection between the flux rope and the external field, cannot be ruled out in the confinement of the eruption (Figure \ref{fig:cartoon}(c)).

\end{enumerate}

\centerline{ACKNOWLEDGMENTS}

We are grateful to Xin Cheng, Chen Xing, Jun Chen, Yang Guo, Jinhan Guo, and Ye Qiu for inspiring discussions. This work was supported by National Key R\&D Program of China under grants 2021YFA1600504 and 2022YFF0503004, and by NSFC under grant 12127901. C.W. is funded by Postgraduate Research \& Practice Innovation Program of Jiangsu Province KYCX23\_0118. F.C. acknowledges the support from the Program for Innovative Talents and Entrepreneurs in Jiangsu. Z.L. is supported by Postgraduate Research \& Practice Innovation Program of Jiangsu Province KYCX22\_0107. This material is based upon work supported by the National Center for Atmospheric Research, which is a major facility sponsored by the National Science Foundation under Cooperative Agreement No. 1852977. The high-performance computing support is provided by Cheyenne (doi:10.5065/D6RX99HX).

\bibliography{manuscript}{}
\bibliographystyle{aasjournal}

\listofchanges
\end{document}